\documentclass[conference]{IEEEtran}
\IEEEoverridecommandlockouts
\usepackage{cite}
\usepackage{amsmath,amssymb,amsfonts}
\usepackage{algorithmic}
\usepackage{graphicx}
\usepackage{textcomp}
\usepackage{xcolor}
\def\BibTeX{{\rm B\kern-.05em{\sc i\kern-.025em b}\kern-.08em
    T\kern-.1667em\lower.7ex\hbox{E}\kern-.125emX}}
    
\usepackage{float}

\usepackage{stfloats}
\usepackage{lipsum}

\usepackage{amsthm}

\usepackage{amssymb}
\usepackage{mathtools}
\usepackage{cases}

\usepackage{booktabs}
\usepackage{multirow}

\usepackage{algorithmic}
\usepackage{algorithm}

\usepackage{amsmath,amsfonts,amsthm,bm} 

\usepackage{comment}

\begin{document}

\title{IRS-aided MIMO Systems over Double-scattering Channels: Impact of Channel Rank Deficiency}

\author{\IEEEauthorblockN{Xin~Zhang\IEEEauthorrefmark{1},
Xianghao~Yu\IEEEauthorrefmark{1},
S.H.~Song\IEEEauthorrefmark{1}, and
Khaled~B.~Letaief\IEEEauthorrefmark{1}\IEEEauthorrefmark{2}}
\IEEEauthorblockA{\IEEEauthorrefmark{1}Dept. of ECE, The Hong Kong University of Science and Technology, Hong Kong}
\IEEEauthorblockA{\IEEEauthorrefmark{2}Peng Cheng Laboratory, Shenzhen 518066, China}
\IEEEauthorblockA{Email:~\IEEEauthorrefmark{1}xzhangfe@connect.ust.hk,~\IEEEauthorrefmark{1}\{eexyu,~eeshsong,~eekhaled\}@ust.hk
}
}

\maketitle

\begin{abstract}
Intelligent reflecting surfaces (IRSs) are promising enablers for next-generation wireless communications due to their reconfigurability and high energy efficiency in improving poor propagation condition of channels, e.g., limited scattering environment. However, most existing works assumed full-rank channels requiring rich scatters, which may not be available in practice. To analyze the impact of rank-deficient channels and mitigate the ensued performance loss, we consider a large-scale IRS-aided MIMO system with statistical channel state information (CSI), where the \textit{double-scattering channel} is adopted to model rank deficiency. 
By leveraging random matrix theory (RMT), we first derive a deterministic approximation (DA) of the ergodic rate with low computational complexity and prove the existence and uniqueness of the DA parameters. Then, we propose an alternating optimization algorithm for maximizing the DA with respect to phase shifts and signal covariance matrices. Numerical results will show that the DA is tight and our proposed method can effectively mitigate the performance loss induced by channel rank deficiency.
\end{abstract}

\section{Introduction}
\label{Introduction}
While the demand for wireless capacity will continue to grow and be met by 5G networks, the emergence of the Internet of Everything (IoE), connecting billions of people and potentially trillions of machines, already calls for the development of 6G~\cite{letaief2019roadmap}. The development of 6G is far from trivial. Many technical challenges must be tackled and stringent requirements must be met that are beyond the reach of current systems, such as extremely high bit rates of up to $1$ Tbps and massive connections reaching at least $10^6$ devices/km\textsuperscript{2}. Recently, intelligent reflecting surfaces (IRSs) have been proposed as a promising and disruptive enabling technology for high-capacity future wireless communications systems, due to its ability to customize wireless propagation environment~\cite{huang2019reconfigurable}. With smartly controlled phase shifters, IRSs can change the end-to-end signal propagation direction and improve coverage with low energy consumption because of its passive nature.


Given that two reflecting channels need to be estimated without active radio frequency chains at IRSs, there are significant challenges and overhead in acquiring instantaneous channel state information (CSI). More importantly, it is impractical to adaptively change the phase shifts according to instantaneous CSI~\cite{zappone2021intelligent}. To address this issue, many works considered the analysis and design of IRS-aided wireless systems with statistical CSI~\cite{kammoun2020asymptotic,zhang2020transmitter,gao2020distributed}. In~\cite{kammoun2020asymptotic} and~\cite{zhang2020transmitter}, the authors first derived an accurate deterministic approximation (DA) of the minimum signal-to-interference-plus-noise-ratio (SINR) and ergodic rate, respectively, by utilizing random matrix theory (RMT). Then, based on the DA, the minimum SINR of a multiuser multiple-input-single-output (MISO) system was maximized in~\cite{kammoun2020asymptotic} by designing the optimal linear precoder, while the phase shifts and the covariance matrix of the transmitted signal were jointly optimized to maximize the ergodic rate for a multiple-input-multiple-output (MIMO) system in~\cite{zhang2020transmitter}. In~\cite{gao2020distributed}, a closed-form expression for the ergodic rate of a MISO system with distributed IRSs was obtained and an efficient phase shift design algorithm was proposed to maximize the ergodic rate. However, all these works assumed full-rank models, which does not reflect the real propagation environment in IRS-aided systems.

In realistic scenarios, IRSs are typically deployed to help wireless transmission with poor propagation conditions, e.g., limited scattering, which leads to a low-rank channel matrix. The rank deficiency not only severely degrades the capacity of MIMO channels but also increases the spatial correlation~\cite{shin2003capacity}. The double-scattering channel was proposed to characterize more realistic channel scenarios
~\cite{gesbert2002outdoor},~\cite{hoydis2011asymptotic}, and was used to model the poor scattering channel in millimeter wave (mmWave) systems~\cite{papazafeiropoulos2017impact}. The rank deficiency, the spatial correlation, and the signal attenuation can be reflected by the number of scatterers, the correlation matrices, and the large-scale fading coefficients in the model, respectively~\cite{van2021uplink}.

In this paper, we consider deploying an IRS to assist a point-to-point MIMO communication system over double-scattering fading channels, where the direct link between the transmitter and the receiver is blocked. 
Our objective is to investigate the impact of the channel rank deficiency and maximize the ergodic rate by optimizing the phase shifts at the IRS and the signal covariance matrix at the transmitter with statistical CSI. To achieve this goal, a DA of the ergodic rate is first derived by capitalizing on the Stieltjes and Shannon transform~\cite{couillet2010deterministic}. The DA is proved to be tight in the asymptotic regime and is also shown to be accurate even in low dimensions by numerical results. Meanwhile, we prove the existence and uniqueness of the DA parameters. We then propose an alternating optimization (AO) algorithm to maximize the DA with respect to the phase shift matrix and signal covariance matrix, which has been shown to be efficient in solving IRS optimization problems~\cite{ma2020low},~\cite{yu2021irs}. The main contributions of this paper include:
\begin{itemize}
  \item[1.] We evaluate and optimize the ergodic rate of IRS-aided communication systems over a double-scattering channel, which is the first attempt in the literature.
  \item[2.] We investigate the impact of rank deficiency on system performance and show that the proposed algorithm can significantly mitigate the influence of rank deficiency.
\end{itemize}

The rest of this paper is organized as follows. In Section~\ref{channel_pro}, we introduce the double-scattering channel model and formulate the problem. In Section~\ref{main_res}, an asymptotic approximation of the ergodic rate is presented. In Section~\ref{opt_sec}, an algorithm is proposed to design the optimal signal covariance and phase shift matrix based on the derived approximation. Numerical results in Section~\ref{simulation} will validate the tightness of the approximation and the effectiveness of the proposed algorithm. Finally, Section~\ref{conclusions} concludes the paper. The notations used throughout the paper are listed in the footnote\footnote{\underline{Notations}. {Throughout this paper, we use boldfaced lowercase letters to represent column vectors and boldfaced uppercase letters to represent matrices. The notation $\mathbb{H}^{N}$ denotes the space of the $N$-dimensional Hermitian matrices. $\mathbb{E} \left[x\right]$ represents the expectation of random variable $x$. Furthermore, ${\rm diag}(\boldsymbol{a})$ represents a square diagonal matrix with the elements of the vector $\boldsymbol{a}$ constituting the main diagonal. The superscript `H' represents the Hermitian transpose, while ${\rm Tr}(\bold{A})$ and $\|\bold{A} \| $ represent the trace and spectral norm of $\bold{A}$, respectively. $\otimes$ denotes the Hadamard product. Function $(x)^{+}=\max\left\{0, x \right\}$. Additionally, $\overset{a.s}{\longrightarrow}$ is used to represent convergence almost surely.}} below.

\section{IRS-aided Communication}
\label{channel_pro}

 \begin{figure}[t]
\centering\includegraphics[width=0.4\textwidth]{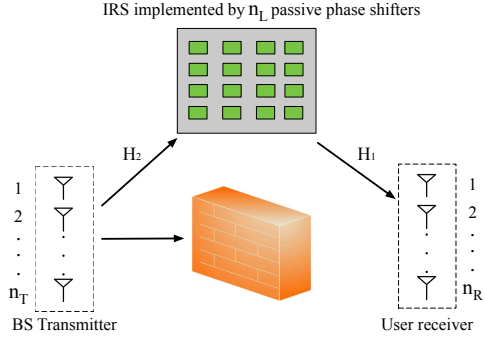}
\caption{An IRS-aided MIMO communication system.}
\label{system_fig}
\vspace{-0.5cm}
\end{figure}

As shown in Fig.~\ref{system_fig}, we consider the downlink of an IRS-aided MIMO communication system, where there is a base station (BS) with $n_{T}$ transmitter antennas and a user with $n_{R}$ receive antennas. The IRS is deployed to establish favorable communication links for the user that would otherwise be blocked. The IRS is assumed to be implemented by $n_{L}$ passive phase shifters. Therefore, the received signal $\bold{y}$ at the receiver is given by
\begin{equation}
\bold{y} = \bold{H}_{1}\bold{\Phi}\bold{H}_{2}\bold{s}+\bold{n},
\end{equation}
where $\bold{s} \in \mathbb{C}^{n_{T}}$ represents the transmitted signal and $\bold{n} \in \mathbb{C}^{n_{R}}$ denotes the additive white Gaussian noise with variance $\sigma^{2}$. $ \bold{H}_{1}  \in \mathbb{C}^{n_{R}\times n_{L}} $ represents the channel matrix from the IRS to user and $ \bold{H}_{2}  \in \mathbb{C}^{n_{L}\times n_{T}} $ represents the channel matrix from the BS to IRS. $\bold{\Phi}=\mathrm{diag}\left(\phi_{1}, \phi_{2},...,  \phi_{n_L}\right)$ with $\phi_{i}=e^{j\theta_{i}}, i=1,2,...,n_L$ represents the phase shifts implemented at the IRS. The normalized ergodic rate of the MIMO channel is given by
\begin{equation}
\label{r_phi_q}
R(\bold{\Phi},\!\bold{Q})\!=\!\frac{1}{n_{R}}\!\mathbb{E} \!\left[\!
\log \det\left(\!\bold{I}\!+\!\frac{1}{\sigma^{2} }\!\bold{H}_{1}\bold{\Phi}\bold{H}_{2} \bold{Q}\bold{H}_{2}^{H}\bold{\Phi}^{H}  \bold{H}_{1}^{H}\!\right)\!\right]\!,\!\!\!
\end{equation}
in bits per second per Hz per antenna, where $\bold{Q}=\mathbb{E}\left[ \bold{s}\bold{s}^H \right]$ represents the covariance matrix  of the transmitted signal and
the expectation is taken over all random matrices, e.g., $\bold{H}_1, \bold{H}_2$.

\subsection{Channel Model}
In this paper, we consider the double-scattering channel~\cite{gesbert2002outdoor}, which is more general and commonly adopted to model the rank deficiency issue in mmWave systems. 
With the double-scattering model, the channel matrix can be written as
\begin{equation}
\label{dc_ch}
\bold{H}_{i}=\bold{R}_{i}^{\frac{1}{2}} \bold{X}_{i} \bold{S}_{i}^{\frac{1}{2}} \bold{Y}_{i}\bold{D}_{i}^{\frac{1}{2}} , i=1,2.
\end{equation}
where $\bold{R}_{i} \in \mathbb{H}^{n_{R_{i}}} $, $\bold{S}_{i} \in \mathbb{H}^{n_{S_{i}}} $ and $\bold{D}_{i} \in \mathbb{H}^{n_{D_{i}}} $ represent the transmit, scatterer, and receive correlation matrices, respectively, and $n_{S_{i}}$ denotes the number of scatters. The rank deficiency is reflected by $n_{S_{i}} < \min\left\{ n_{R_{i}},n_{D_{i}} \right\}$. Correspondingly, we have $n_{R}=n_{R_1}, n_{L}=n_{D_1}=n_{R_2}$ and $n_{D_2}=n_{T}$. $\bold{X}_{i} \in \mathbb{C}^{n_{R_{i}}\times n_{S_{i}}}$ and $\bold{Y}_{i} \in \mathbb{C}^{n_{S_{i}}\times n_{D_{i}}} $ are independent and their entries $X_{i,j}$ and $Y_{i,j}$ are independent and identically distributed (i.i.d.) complex zero-mean Gaussian random variables with variance $\frac{1}{n_{S}}$ and $\frac{1}{n_{D}}$, respectively.

\subsection{Problem Formulation}
Our objective is to maximize the normalized ergodic rate with respect to the phase shift matrix $\bold{\Phi}$ and signal covariance matrix $\bold{Q}$ under energy constraints, which can be formulated as
\begin{equation}
\begin{split}
\mathcal{P}1:~&\max\limits_{\bold{Q}, \bold{\Phi}} ~{R\left(\bold{\Phi},\bold{Q} \right)},\hfill\\
~\mathrm{s.t.}~&\mathrm{Tr} \left( \bold{Q}\right) \le n_{T}P, ~\bold{Q} \succeq 0, \\ 
 &\bold{\Phi}=\mathrm{diag} \left\{\phi_1,...,\phi_{n_{L}} \right\}, |\phi_i|=1, i=1,...,n_{L}.
\end{split}
\end{equation}
The challenge in solving $\mathcal{P}1$ comes from two aspects. First, the objective function is an expectation form over four random matrices, which is usually evaluated by numerical methods with a huge amount of computation payload. Second, the feasible set of $\mathcal{P}1$ is highly non-convex due to the uni-modular constraints on phase shifts. 

\section{An Asymptotic Approximation for Ergodic Rate with Double-scattering Channel}
\label{main_res}
Since there are two random matrices in~(\ref{dc_ch}), it is challenging to evaluate~(\ref{r_phi_q}), not only in closed forms but also via numerical methods. Therefore, an accurate and computationally-efficient approximation is desired. In this section, we leverage random matrix theory to derive the DA of the normalized ergodic rate. First of all, we present two assumptions, based on which the DA is developed.

\textbf{Assumption 1}: By letting ${\alpha}_{R_i,S_j}=\frac{n_{R_i}}{n_{S_j}}, {\alpha}_{S_i,D_j}=\frac{n_{S_i}}{n_{D_j}}$, then $0 < \lim\inf  {\alpha}_{R_i,S_j} < \infty, 0 < \lim\inf {\alpha}_{S_i,D_j} < \infty, i,j=1,2$.\\
This assumption indicates that the scale of the IRS and the numbers of antennas and scatters grow in the same order.

\textbf{Assumption 2}: $\lim\sup  \| \bold{R}_{i} \|< \infty$, $\lim\sup  \| \bold{S}_{i} \|< \infty$ and $\lim\sup  \| \bold{T}_{i} \|< \infty$, for $i=1,2$, where

\begin{equation}
\label{transform_t}
\left\{
\begin{aligned}
\bold{T}_{1}&=\bold{\Phi}^{H}  \bold{D}_{1}  \bold{\Phi},\\
\bold{T}_{2} &= \bold{D}^{\frac{1}{2}}_{2}\bold{Q} \bold{D}^{\frac{1}{2}}_{2}.
\end{aligned}
\right.
\end{equation}
The normalized ergodic rate in~(\ref{r_phi_q}) can be written as 
\begin{equation}
R(\bold{\Phi},\bold{Q})=\frac{1}{n_{R_1}}\mathbb{E} \left[
\log \det\left(\bold{I}+\frac{1}{\sigma^{2} }\bold{B} \right)\right] ,
\end{equation} 
where $\bold{B}=\bold{H}'\bold{H}'^{H}$, $\bold{H}'=\bold{H}'_1 \bold{H}'_2$ and $\bold{H}'_i=\bold{R}_{i}^{\frac{1}{2}} \bold{X}_{i} \bold{S}_{i}^{\frac{1}{2}} \bold{Y}_{i}\bold{T}_{i}^{\frac{1}{2}}$. The matrix $\bold{\Phi}^{H}$ can be omitted because $\bold{Y}_{i}\bold{\Phi}^{H}$ has the same statistical properties as $\bold{Y}_{i}$ due to the unitary-invariant attribute of the Gaussian distribution. According to Shannon transform~\cite{hachem2007deterministic}, we have
\begin{equation}
\begin{aligned}
\label{r_std}
R(\bold{\Phi},\bold{Q})
= \int_{\frac{1}{\sigma^2}}^{\infty} \frac{1}{z} -  \mathbb{E} \left[ m_{\bold{B}}(z) \right]dz=R(\sigma^2),
\end{aligned}
\end{equation}
where 
\begin{equation}
m_{\bold{B}}(z)  = \frac{1}{n_{R_1}} \mathrm{Tr} \left(  \left( z \bold{I} + \bold{B} \right)^{-1} \right)
\end{equation}
is the Stieltjes transform of the limiting spectral distribution (LSD) of matrix $\bold{B}$ for $z \in \mathbb{C}-\mathbb{R}^{-} $. 

Hence, it is essential to derive an asymptotic approximation of $\mathbb{E} \left[ m_{\bold{B}}(z) \right]$ and we have the following theorem regarding the approximation.

{\bfseries Theorem~1}: The DA of $\mathbb{E} \left[ m_{\bold{B}}(z)\right] $ is given by
\begin{equation}
\mathbb{E} \left[ m_{\bold{B}}(z) \right] \xrightarrow{n_{R_1}   \rightarrow \infty}
\frac{1}{n_{R_{1}}}  \mathrm{Tr} \left( \left( z \bold{I}+  h_{2}h_{3}h_{4}h_{5}\bold{R}_{1} \right)^{-1}\right),
\end{equation}
where $h_i,i=1,2,...,5$, satisfy the following system of equations with a unique positive solution
  \begin{subequations}\label{de_group}
    \begin{alignat}{4}
      & h_1=\frac{1}{n_{R_1}} \mathrm{Tr}\left( \bold{R}_{1} \left(z\bold{I}+h_{2}h_{3}h_{4}h_{5} \bold{R}_{1}  \right)^{-1}\right),
       \label{eqt1}  \\
     &h_2=\frac{1}{n_{S_1}} \mathrm{Tr}\left( \bold{S}_{1} \left( \bold{I}+  \alpha_{R_1,S_1} h_{1}h_{3}h_{4}h_{5} \bold{S}_{1}  \right)^{-1}\right),  \label{eqt2}  \\
    &h_3=\frac{1}{n_{T_1}} \mathrm{Tr}\left(\! \bold{T}_{1}^{\frac{1}{2}} \bold{R}_{2}\bold{T}_{1}^{\frac{1}{2}}\!\! \left(\bold{I}\!+\!\alpha_{R_1, T_1} h_{1}h_{2}h_{4}h_{5} \bold{T}_{1}^{\frac{1}{2}} \bold{R}_{2}\bold{T}_{1}^{\frac{1}{2}}  \right)^{-1}\!\right)\!\!,
\label{eqt3}\\
& h_4=\frac{1}{n_{S_2}} \mathrm{Tr}\left( \bold{S}_{2} \left( \bold{I}+   \alpha_{R_1, S_2}h_{1}h_{2}h_{3} h_{5} \bold{S}_{2}  \right)^{-1}\right), \label{eqt4}\\
& h_5= \frac{1}{n_{T_2}}  \mathrm{Tr}\left( \bold{T}_2\left( \bold{I}+ \alpha_{R_1,T_2} h_{1} h_{2} h_{3}h_{4}  \bold{T}_2 \right)^{-1}\right). \label{eqt5}
    \end{alignat}
  \end{subequations}

\begin{IEEEproof}
Please refer to Appendix~\ref{append}.
\end{IEEEproof}

Note that $\bold{R}_i, \bold{S}_i, \bold{T}_i,i=1,2$ can be arbitrary positive semi-definite matrices. Based on {\bfseries Theorem~1} and~(\ref{r_std}), we further have the following result for the DA of $R\left(\bold{\Phi},\bold{Q} \right)$.

{\bfseries Theorem~2}: The DA of $R\left(\bold{\Phi},\bold{Q} \right)$ is given by
\begin{equation}
R\left(\bold{\Phi},\bold{Q} \right) \xrightarrow{n_{R_1}   \rightarrow \infty}  \overline{R}\left( \bold{\Phi},\bold{Q}\right),
\end{equation}
where $\overline{R}\left( \bold{\Phi},\bold{Q}\right)$ is given in~(\ref{de_obj}) at the bottom of the next page.

\begin{figure*}[b]
\normalsize
\vspace{-0.3cm}
\hrulefill
\begin{equation}
\begin{aligned}
\label{de_obj}
 \overline{R}\left(\bold{\Phi},\bold{Q}  \right) =  
&  \frac{1}{n_{R_{1}}} \Big[  \mathrm{log det}\left( \bold{I} + \frac{1}{\sigma^{2}}h_{2}h_{3}h_{4}h_{5} \bold{R}_{1} \right)+ 
\mathrm{logdet}\left( \bold{I} + { \alpha}_{R_1,S_1} h_{1}h_{3}h_{4}h_{5} \bold{S}_{1} \right)
+
 \mathrm{log det}\left( \bold{I} +  {\alpha}_{R_1, S_2}h_{1}h_{2}h_{3} h_{5} \bold{S}_{2}   \right)
\\
+&\mathrm{log det}\left( \bold{I} +  {\alpha}_{R_1, D_1} h_{1}h_{2}h_{4}h_{5} \bold{D}_{1}\bold{\Phi}\bold{R}_{2} \bold{\Phi}^{H} \right)
+\mathrm{log det}\left( \bold{I}+ { \alpha}_{R_1,D_2} h_{1} h_{2} h_{3}h_{4}   \bold{D}^{\frac{1}{2}}_{2}\bold{Q} \bold{D}^{\frac{1}{2}}_{2}\right)\Big]
-4 h_{1}h_{2}h_{3}h_{4}h_{5}
\end{aligned}
\end{equation}
\end{figure*}

\begin{IEEEproof}
Please refer to Appendix~\ref{append2}.
\end{IEEEproof}

\section{Optimization of the Signal Covariance Matrix and Phase Shift Matrix}
\label{opt_sec}
In this section, based on the derived DA in~(\ref{de_obj}), we propose an AO algorithm to jointly optimize the phase shift matrix $\bold{\Phi}$ and signal covariance matrix $\bold{Q}$ to maximize the ergodic rate.

\subsection{Problem Reformulation}
 We adopt the DA form derived in~(\ref{de_obj}) as an asymptotic approximation of the objective function in $\mathcal{P}1$. Hence, we re-state the DA optimization problem as follows:
\begin{equation}
\begin{aligned}
\mathcal{P}2:~&\max\limits_{\bold{Q}, \bold{\Phi}} ~\overline{R}\left(\bold{\Phi},\bold{Q} \right)~,\\
 ~\mathrm{s.t.}&~ \mathrm{Tr} \left(\bold{Q}\right) \le n_{D_{2}}P, ~\bold{Q} \succeq 0, \\ 
 &~\bold{\Phi}=\mathrm{diag} \left\{\phi_1,...,\phi_{n_{L}} \right\}, |\phi_i|=1, i=1,...,n_{L}.
\end{aligned}
\end{equation}
Due to the non-convexity of $\mathcal{P}2$, we resort to the AO technique to jointly optimize $\bold{\Phi}$ and $\bold{Q}$. Specifically, an $(n_L+1)$-block AO approach is developed, where the correlation matrix of the transmitted signal and the phase shifts of the IRS are optimized alternately.

\subsection{Signal Covariance Matrix Optimization}
\label{sig_cov}
With a given $\bold{\Phi}$, $\mathcal{P}2$ is the maximization of a concave function with convex constraints and it becomes
\begin{equation}
\begin{aligned}
\mathcal{P}3:~&\max\limits_{\bold{Q}} \frac{1}{n_{R_1}}\mathrm{log ~ det}( \bold{I}+  {\alpha}_{R_1,D_2} h_{1} h_{2} h_{3}h_{4}  \bold{D}_{2}\bold{Q}  )\\
~\mathrm{s.t.}&~\mathrm{Tr} \left(\bold{Q} \right)\le n_{D_{2}}P.
\end{aligned}
\end{equation}
In fact, $\mathcal{P}$3 can be directly solved by the standard water-filling method, which is also adopted in~\cite{li2008transmitter} to design optimum transmission direction. The optimal solution is given by,
\begin{equation}
\label{opt_q}
\hat{\bold{Q}}=\bold{U} \bold{\Lambda}_{Q}\bold{U}^{H},
\end{equation}
where the unitary matrix $\bold{U}$ and the diagonal matrix $\bold{\Lambda}_{D_{2}}$ satisfy $\bold{D}_{2}=\bold{U} \bold{\Lambda}_{D_{2}}\bold{U}^{H}$. The diagonal matrix $\bold{\Lambda}_{Q}=(\frac{1}{\mu}\bold{I}-\frac{1}{{\alpha}_{R_1,D_2} h_{1} h_{2} h_{3}h_{4} }\bold{\Lambda}^{-1}_{D_{2}})^{+}$, and $\mu$ is determined by the constraint $\mathrm{Tr}\left( \bold{Q}\right) = \mathrm{Tr} \left(\bold{\bold{\Lambda}_{Q}} \right)= n_{D_{2}}P$. 

\subsection{Phase Shift Matrix Optimization}
\label{pha_opt}
It can be observed from~(\ref{de_group}) and~(\ref{transform_t}) that all $h_{i}$'s are related to the phase shifts. 
By fixing $\bold{Q}$, $\mathcal{P}2$ is transformed to the following sub-problem with respect to $\bold{\Phi}$
\begin{equation}
\begin{aligned}
\mathcal{P}4:&~\max\limits_{\bold{\Phi}}~\overline{R}\left(\bold{\Phi},\bold{Q} \right)=\max\limits_{\bold{\Phi}}~G(\bm{\theta}),
\\
 \mathrm{s.t}.& ~\bold{\Phi}=\mathrm{diag} \left\{\phi_1,...,\phi_{n_{L}} \right\}=\mathrm{diag} \left(\exp\left({ \jmath \bm{\theta}}\right)\right),
 \\
& ~ \theta_i  \in [0,2\pi), i=1,...,n_{L}.
\end{aligned}
\end{equation}


Due to the uni-modular constraint for each phase shift, $\mathcal{P}4$ is a non-convex problem with regard to the reflecting matrix $\bold{\Phi}$. However, we can find a  suboptimal solution by resorting to the gradient method. First, we determine the gradient of $\overline{R}\left(\bold{\Phi},\bold{Q} \right)$ with respect to $\phi_{i}$'s as
\begin{equation}
\label{dereva}
\begin{aligned}
\frac{G(\bm{\theta})}{\partial \theta_{i}}=\frac{ {\alpha}_{R_1, D_1} h_{1}h_{2}h_{4}h_{5} }{n_{R_{1}}}\mathrm{Tr} [\bold{D}_{1}(\bold{R}_{2}\otimes \bold{F} )
\\
 \times\left( \bold{I} +  {\alpha}_{R_1, D_1} h_{1}h_{2}h_{4}h_{5} \bold{D}_{1}\bold{\Phi}\bold{R}_{2} \bold{\Phi}^{H} \right)^{-1}].
\end{aligned}
\end{equation}
In~(\ref{dereva}), $\bold{F}\in \mathbb{C}^{n_L\times n_L}$ is given by
\begin{equation}
\left[\bold{F}\right]_{p,q}=\left\{
\begin{aligned}
& \jmath e^{\jmath (\theta_{i}-\theta_{q})} ,  p = i, \\
& -\jmath e^{\jmath (\theta_{p}-\theta_{i})} ,  q = i, \\
&0,  otherwise. \\
\end{aligned}
\right.
\end{equation}
Here, we use the backtrack line search method~\cite{boyd2004convex} to find the step size $\gamma$ such that
\begin{equation}
\label{grad_up}
G\left(\bm{\theta}+\gamma\nabla G\left(\bm{\theta} \right) \right) \ge  G(\bm{\theta})+
c\gamma \|\nabla G(\bm{\theta})\|,
\end{equation}
where $\nabla G(\bm{\theta})=\left(\frac{G(\bm{\theta})}{\partial \theta_{1}},\frac{G(\bm{\theta})}{\partial \theta_{2}},...,\frac{G(\bm{\theta})}{\partial \theta_{n_{L}}}\right)^{T}$ and $0<c<1$ is a constant.

\subsection{AO Algorithm}
\begin{algorithm} 
\caption{Alternating Optimization on Phase Shift Matrix $\bold{\Phi}$ and Signal Covariance Matrix $\bold{Q}$  } 
\label{alg:1} 
\begin{algorithmic}[1] 
\REQUIRE  $\bm{\theta}^{\left(0 \right)}, \bold{Q }^{\left(0 \right)}$, and set $t=0$.
\REPEAT
\STATE Obtain optimal $\bold{Q}^{(t+1)}$ with given $\bm{\theta}^{(t)}$ by~(\ref{opt_q}).

\STATE Compute $h_{i},i=1,...,5$ according to~(\ref{de_group}) with given $\bm{\theta}^{(t)}$ and $\bold{Q}^{(t+1)}$.
	\STATE Compute $\nabla G\left(\bm{\theta}^{(t)} \right)$ by~(\ref{dereva}).
	\STATE Find the step size $\gamma$ by backtrack line search.
	\STATE $\bm{\theta}^{(t+1)}= \bm{\bold{\theta}}^{(t)}+\gamma  G\left(\bm{\theta}^{(t)} \right)$.
\STATE $t \leftarrow  t+1$
\UNTIL Convergence.
\ENSURE  $\bm{\theta}^{t}, \bold{Q}^{t}$.
\end{algorithmic}
\end{algorithm}

Based on the results in Sections~\ref{sig_cov} and~\ref{pha_opt},  the overall AO algorithm is summarized in \textbf{Algorithm~\ref{alg:1}}. It is worth noting that the solution $\bold{Q}$ in $\mathcal{P}3$ is optimal for its concavity and the $\overline{R}$ in $\mathcal{P}4$ will converge as it is monotonically increasing. Therefore, the AO algorithm is guaranteed to converge.

\section{Numerical Results}
\label{simulation}

In this section, numerical results are provided to validate the effectiveness of the proposed methods. For a double-scattering channel\cite{hoydis2011asymptotic}, the $(l,m)$-th entry of the correlation matrix is given by
\begin{equation}
\left[\bold{C}(\phi,N,d)\right]_{l,m}\!\!=\!\frac{1}{N}\!\!\sum_{n=\frac{1-N}{2}}^{\frac{N-1}{2}}\!\!\exp\left(\jmath 2\pi  d(l-m) \mathrm{sin}\left(\frac{n\phi}{1-N}\right)\! \right)\!,
\end{equation}
where $\phi$ represents the angular spread of the signals, $d$ is the antenna spacings, and $N_s$ is the number of the scatters. Thus, we have $\bold{R}=\bold{C}(\phi_r,n_{s},d_r)$, $\bold{S}=\bold{C}(\phi_s,n_{s}, d_{s})$, and $\bold{D}=\bold{C}(\phi_{t},n_{s},d_t)$ in~(\ref{dc_ch}).
The key parameters are set as: $\phi_{R_i}=\phi_{T_i}=\frac{\pi}{7}$, $\phi_{S_i}=\frac{\pi}{16}$, $d_{t_i}=d_{r_i}=d_{s_i}=25$.

\begin{figure}[H]
\centering\includegraphics[width=0.4\textwidth]{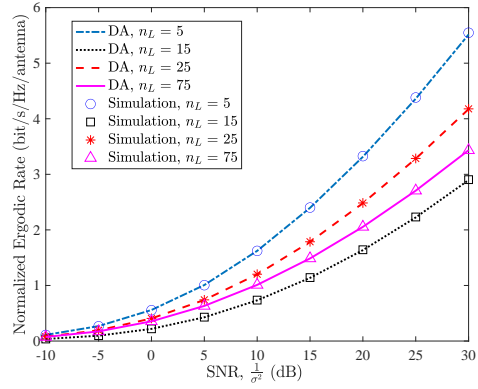}
\vspace{-0.2cm}
\caption{Normalized ergodic rate versus SNR.}
\label{fig:1}
\vspace{-0.2cm}
\end{figure}

In Fig.~\ref{fig:1}, we show the comparison between the DA in~(\ref{de_obj}) and Monto-Carlo simulation results. The parameters are set as $\bold{Q}^{(0)}=\bold{I}$, $\bold{\Phi}^{(0)}=\mathrm{diag}(e^{\frac{2\pi \jmath l}{n_{L}}}),~l=0,1,...,n_{L}-1$, $n_{R_{1}}= n_{D_{2}}= n_{S_{1}}=n_{S_{2}}=5$, $n_{R_{i}}=n_{S_{i}}=n_{T_{i}}=\left\{5,15, 25, 75\right\}, i=1,2$. The number of realizations in Monto-Carlo simulation is 2000. It can be observed from Fig.~\ref{fig:1} that the derived DA can approximate the normalized ergodic rate accurately and the approximation remains tight even for the low dimension case ($n_{L}=5$).

\begin{figure}[t!]
\vspace{-0.3cm}
\centering\includegraphics[width=0.4\textwidth]{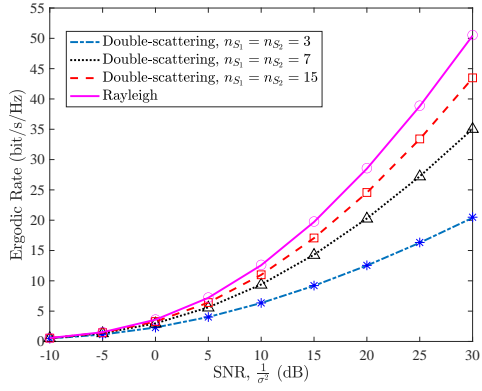}
\vspace{-0.2cm}
\caption{Effect of rank deficiency.}
\label{fig_rank}
\end{figure}

To illustrate the performance loss caused by rank-deficiency, we compare the ergodic rate of double-scattering channel with different ranks and that of the full-rank Rayleigh channel in Fig.~\ref{fig_rank}. 
The transmit and receive correlation matrices for the Rayleigh channel are identical to those for the double-scattering channel, i.e., $\bold{R}_{i}$ and $\bold{T}_{i}, i=1,2$, which are full-rank. Simulation results are also included and denoted with markers. The parameters are set as $n_{R_1}=n_{T_1}=n_{R_2}=n_{T_2}=15$ and $n_{S_1}=n_{S_2}=3,7,15$. We observe that the ergodic rate degrades rapidly as $n_{S_1}$ and $n_{S_2}$ decreases, which corresponds to a more severe rank deficiency. 

\begin{figure}[t!]
\centering\includegraphics[width=0.4\textwidth]{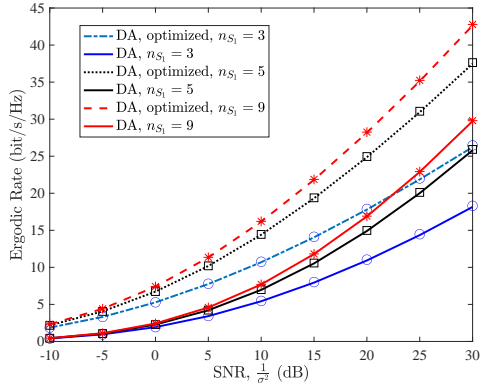}
\vspace{-0.2cm}
\caption{Optimized ergodic rate.}\label{fig:4}
\vspace{-0.4cm}
\end{figure}

In Fig.~\ref{fig:4}, the effectiveness of the proposed optimization algorithm is illustrated when $n_{R_1}=n_{T_1}=n_{R_2}=n_{T_2}=9$ and $n_{S_1}=n_{S_2}=3, 5, 9$ with $c=0.0005$. Simulation results are also included and denoted with markers. By applying the proposed AO algorithm, the ergodic rate loss caused by rank-deficient channels can be efficiently compensated. For example, the optimized rate of low-rank channel ($n_{S_1}=3$) outperforms the full-rank case ($n_{S_1}=9$) when $\mathrm{SNR}$ is lower than $20~\mathrm{dB}$.

\section{Conclusions}
\label{conclusions}

In this paper, we investigated the impact of channel rank-deficiency on IRS-aided systems by considering the double-scattering channel. Given the statistical CSI, we maximized the ergodic rate by optimizing the transmit signal covariance and phase shift matrices. For that purpose, we first derived a DA of the ergodic rate, which is accurate and computationally efficient. Then, an AO algorithm was proposed to obtain the optimal transmit covariance matrix and phase shifts. Numerical results demonstrated not only the accuracy of the evaluation but also the effectiveness of our proposed algorithm in compensating the ergodic rate loss due to the channel rank-deficiency.

    \begin{appendices}
      \section{prove of theorem 1  }
  \label{append}
 
 
\begin{IEEEproof}
The proof includes two parts. First we derive the DA form of $\mathbb{E}\left[m_{\bold{B}}(z)\right]$ through an iterative approach. Then, we show the existence and uniqueness of the solution of~(\ref{de_group}) using the \textit{standard interference function} theory~\cite{yates1995framework}.

Here we use an iterative approach to erase the randomness of each $\bold{X}_{i}$ and $\bold{Y}_{i}$, $i=1,2, j=1,2$.
Let $\widetilde{\bold{T}}=\bold{S}_{1}^{\frac{1}{2}}\bold{Y}_{1}\bold{T}_{1}^{\frac{1}{2}}\bold{R}_{2}^{\frac{1}{2}}\bold{X}_{2}\bold{S}_{2}^{\frac{1}{2}}\bold{Y}_{2}\bold{T}_{2}\bold{S}_{2}^{\frac{1}{2}}\bold{Y}_{2}^{H}\bold{R}_{2}^{\frac{1}{2}}\bold{Y}_{1}^{H}\bold{S}_{1}^{\frac{1}{2}}$ and consider $\widetilde{\bold{T}}$ as a deterministic matrix. Under our assumptions in section~\ref{main_res}, by theorem 1 in~\cite{zhang2013capacity}, we have:
\vspace{-0.2cm}
\begin{equation}
\small
  m_{\bold{B}|\widetilde{\bold{T}}}(z) \overset{a.s.}{\longrightarrow} \frac{1}{n_{R_{1}}} \mathrm{Tr}\left(\bold{\Theta}_1(z)\right),
\end{equation}
\vspace{-0.7cm}

\begin{equation}
\small
\bold{\Theta}_1(z)=\left( z \bold{I}+  \widetilde{\bold{\beta}_1}\bold{R}_{1} \right)^{-1},
\end{equation}
\vspace{-0.8cm}

\begin{equation}
\small
\beta_1=\frac{1}{n_{R_1}} \mathrm{Tr}\left(\bold{R}_{1} \bold{\Theta}_1 (z)\right),
\end{equation}
\vspace{-0.3cm}
\begin{equation}
\small
\widetilde{\beta_1}=\frac{1}{n_{S_1}}  \mathrm{Tr}\left(\widetilde{\bold{T}}\left( \bold{I}+ \beta_1 \alpha_{R_1,S_1} \widetilde{\bold{T}} \right)^{-1}\right).
\end{equation}
\vspace{-0.4cm}

Since the remaining randomness lies in matrix $\widetilde{\bold{T}}$, by letting $\widetilde{\widetilde{\bold{T}}}=\bold{T}_{1}^{\frac{1}{2}}\bold{R}_{2}^{\frac{1}{2}}\bold{X}_{2}\bold{S}_{2}^{\frac{1}{2}}\bold{Y}_{2}\bold{T}_{2}\bold{S}_{2}^{\frac{1}{2}}\bold{Y}_{2}^{H}\bold{R}_{2}^{\frac{1}{2}}$ and referring to Theorem 1 in~\cite{zhang2013capacity}, we have:
\begin{equation}
\small
  m_{ \widetilde{\bold{T}  } |\widetilde{\widetilde{\bold{T}}}}(z)  \overset{a.s.}{\longrightarrow}  \frac{1}{n_{S_1}} \mathrm{Tr}\left(\bold{\Theta}_2\left( z\right)\right) ,
\end{equation}

\vspace{-0.3cm}

\begin{equation}
\small
\bold{\Theta}_2\left( z\right)=\left( z \bold{I}+  \widetilde{\bold{\beta}}_2\bold{S}_1 \right)^{-1},
\end{equation}
\begin{equation}
\small
\beta_2=\frac{1}{n_{S_1}} \mathrm{Tr}\left(\bold{S}_1 \bold{\Theta}_{2}(z)\right),
\vspace{-0.2cm}
\end{equation}
\begin{equation}
\small
\widetilde{\beta_2}=\frac{1}{n_{T_1}}  \mathrm{Tr}\left( \widetilde{\widetilde{\bold{T}}}\left( \bold{I}+ \beta_2 \alpha_{S_1,T_1} \widetilde{\widetilde{\bold{T}}} \right)^{-1}\right).
\vspace{-0.2cm}
\end{equation}
Then, we further derive $\widetilde{\beta_1}$ as
\begin{equation}
\small
\label{firstiter}
\nonumber
\begin{split}
\widetilde{\beta_1}=& \frac{1}{n_{S_{1}}} \frac{1}{\beta_1\alpha_{R_1,S_1}}  \mathrm{Tr} \left( \bold{I} -  \left( \bold{I}+ \beta_1 \alpha_{R_1,S_1} \widetilde{\bold{T}} \right)^{-1} \right)\\
=& \! \frac{1}{\beta_1\alpha_{R_1,S_1}}\!\left( \!\frac{1}{n_{S_1}}  \mathrm{Tr} \left( \bold{I}\right) \!- \! \frac{1}{\beta_1\alpha_{R_1,S_1}}m_{ \widetilde{\bold{T}  }|\widetilde{\widetilde{\bold{T}}}}\left(\frac{1}{\beta_1\alpha_{R_1,S_1}}\right)\right) \\
\overset{a.s.}{\rightarrow}&\!  \frac{1}{\beta_1\alpha_{R_1,S_1}}\! \left(\! \frac{1}{n_{S_1}}\! \mathrm{Tr} \!\left(\! \bold{I} \right) \!- \! \frac{1}{\beta_1\alpha_{R_1,S_1}}\frac{1}{n_{S_1}} \mathrm{Tr}\left(\bold{\Theta}_2 \left(\frac{1}{\beta_1\alpha_{R_1,S_1}} \!\right) \!\right) \!  \right)\\
\end{split}
\vspace{-0.5cm}
\end{equation}
\begin{equation}
\small
= \frac{\widetilde{\beta_2}}{\beta_1\alpha_{R_1,S_1}n_{S_1}} \mathrm{Tr} \left( \bold{S}_1 \bold{\Theta}_2\left(\frac{1}{\beta_1\alpha_{R_1,S_1}} \right)\right)  =\frac{\widetilde{\beta_2}\beta_2}{\beta_1\alpha_{R_1,S_1}}. 
\end{equation}

With the following manipulations,
\begin{equation}
\small
f_1=\beta_1, f_2=\frac{\beta_2}{\alpha_{R_1,S_1}\beta_1}, f_3=\widetilde{\beta_2},
\vspace{-0.2cm}
\end{equation}
we can obtain
\begin{subequations}\label{mid_group}
\small
\begin{equation}
f_1=\frac{1}{n_{R_1}} \mathrm{Tr}\left(\bold{R}_{1} \left(z\bold{I}+f_{2}f_{3} \bold{R}_{1}  \right)^{-1}\right),
\vspace{-0.2cm}
\end{equation}

\begin{equation}
\small
f_2=\frac{1}{n_{S_1}} \mathrm{Tr}\left(\bold{S}_{1} \left( \bold{I}+  \alpha_{R_1,S_1} f_{1}f_{3} \bold{S}_{1}  \right)^{-1}\right),
\vspace{-0.2cm}
\end{equation}

\begin{equation}
\small
\label{itef3}
f_3= \frac{1}{n_{T_1}}  \mathrm{Tr} \left(\widetilde{\widetilde{\bold{T}}}\left( \bold{I}+ \alpha_{R_1,T_1}  f_{1}f_{2} \widetilde{\widetilde{\bold{T}}} \right)^{-1}\right),
\vspace{-0.2cm}
\end{equation}
\end{subequations}
and
\begin{equation}
\small
  m_{\bold{B}|\widetilde{\widetilde{\bold{T}}}}(z) \overset{a.s.}{\longrightarrow} \frac{1}{n_{R_1}}  \mathrm{Tr}  \left(\left(z\bold{I}+f_{2}f_{3} \bold{R}_{1}  \right)^{-1}\right),
\vspace{-0.2cm}
\end{equation}
whose proof is similar to Theorem 1 in~\cite{zhang2020transmitter}. Following the same approach, we have
\vspace{-0.2cm}
\begin{subequations}
\small
\begin{equation}
 m_{\bold{B}|\widetilde{\widetilde{\bold{T}}}}(z) \overset{a.s.}{\longrightarrow} \frac{1}{n_{T_1}}  \mathrm{Tr} \left(\left(z\bold{I}+g_{2}g_{3} \bold{T}_{1}^{\frac{1}{2}} \bold{R}_{2}\bold{T}_{1}^{\frac{1}{2}} \right)^{-1}\right),
\vspace{-0.2cm}
\end{equation}
and the following equations,
\vspace{-0.2cm}
\begin{equation}
\small
g_1=\frac{1}{n_{T_1}} \mathrm{Tr}\left(\bold{T}_{1}^{\frac{1}{2}} \bold{R}_{2}\bold{T}_{1}^{\frac{1}{2}} \left(z\bold{I}+g_{2}g_{3} \bold{T}_{1}^{\frac{1}{2}} \bold{R}_{2}\bold{T}_{1}^{\frac{1}{2}}  \right)^{-1}\right),
\vspace{-0.2cm}
\end{equation}
\begin{equation}
g_2=\frac{1}{n_{S_2}} \mathrm{Tr} \left(\bold{S}_{2} \left( \bold{I}+  \alpha_{T_1,S_2} g_{1}g_{3} \bold{S}_{2}  \right)^{-1}\right),
\vspace{-0.2cm}
\end{equation}
\begin{equation}
g_3= \frac{1}{n_{T_2}}  \mathrm{Tr}\left( \bold{T}_2\left( \bold{I}+ \alpha_{T_1,T_2} g_{1}g_{2}  \bold{T}_2 \right)^{-1}\right).
\vspace{-0.1cm}
\end{equation}
\end{subequations}

From~(\ref{itef3}), we have the following relation between $f_i$ and $g_i$, $i=1,2,3$, by using the techniques same as~(\ref{firstiter}) 
\begin{equation}
\small
f_3 \overset{a.s.}{\longrightarrow} \frac{1}{n_{T_1}\alpha_{R_1,T_1}f_{1}f_{2}}g_{1}g_{2}g_{3}.
\vspace{-0.1cm}
\end{equation}

To get a universal result, we replace $f_{i}$ with $h_{i}$, $i=1,2$, and 
let $h_3=\frac{g_1}{\alpha_{R_1,T_1}f_{1}f_{2}}=\frac{g_1}{\alpha_{R_1,T_1}h_{1}h_{2}}$, $g_2=h_4$, $g_3=h_5$ to obtain equations~(\ref{eqt3}) to~(\ref{eqt5}). Then, we integrate $m_{\bold{B}}(z)$ over the subspace with measure $1$ of the whole probability space to obtain
\vspace{-0.2cm}
\begin{equation}
\small
\mathbb{E} \left[m_{\bold{B}}(z)\right]  \xrightarrow{n_{R_1}   \rightarrow \infty}    \frac{1}{n_{R_{1}}} \mathrm{Tr} \left(\left( z \bold{I}+  {h_2}{h_3}{h_4}{h_5}\bold{R}_{1} \right)^{-1} \right).
\vspace{-0.2cm}
\end{equation}

Next, we will use the \textit{standard interference function} theory~\cite{yates1995framework} to show that the system of equations in~(\ref{de_group}) has a unique solution. Define the function $e(h_1)$ as
\vspace{-0.2cm}
\begin{equation}
\small
e(h_1):=\frac{1}{n_{R_1}} \mathrm{Tr}\left(\bold{R}_{1} \left(z\bold{I}+h_{2}h_{3}h_{4}h_{5} \bold{R}_{1}  \right)^{-1}\right),
\vspace{-0.2cm}
\end{equation}
where $h_2,h_3,h_4,h_5$ satisfy equations~(\ref{eqt2}) to~(\ref{eqt5}). First we will show that $e(h_1)$ is a \textit{standard interference function} according to its definition in~\cite{yates1995framework}. Specifically, we have the following 3 steps:

i). Positivity: It can be readily checked.

ii). Monotonicity: This step is to show that $e(h_1)$ is an increasing function. Assume that $h_1 > h'_1$,
From~(\ref{eqt2}), we have
\vspace{-0.2cm}
\begin{equation}
\small
\begin{aligned}
1&=g(h_1,h_2,h_3,h_4,h_5) \\    &=\frac{1}{n_{S_1}} \mathrm{Tr}\left(\bold{S}_{1} \left( h_2\bold{I}+  \alpha_{R_1,S_1} h_{1}h_{2}h_{3}h_{4}h_{5} \bold{S}_{1}  \right)^{-1}\right),
\end{aligned}
\vspace{-0.2cm}
\end{equation}
where $g(h_1,h_2,h_3,h_4,h_5)$ is decreasing with $h_2$ and $M:=h_{1}h_{2}h_{3}h_{4}h_{5}$. If $h_2>h'_2$, there must be an $M<M'=h'_{1}h'_{2}h'_{3}h'_{4}h'_{5}$, which implies that $h_3>h'_3, h_4>h'_4, h_5>h'_5$ from~(\ref{eqt3}) to~(\ref{eqt5}). Therefore, there must be $h_1<h'_1$, which is conflict with our assumption. Hence, there should be $h_2<h'_2, h_3<h'_3, h_4<h'_4, h_5<h'_5$ and $M>M'$, from which we conclude that $e(h_1)>e(h'_1)$. 

iii). Scalability: For $\alpha>1$, using the matrix equation $A^{-1}-B^{-1}=A^{-1}\left(B-A \right)B^{-1}$, we have
\vspace{-0.2cm}
\begin{equation}
\nonumber
\small
\begin{aligned}
&\alpha e(h_1) - e(\alpha h_1)= \frac{1}{n_{R_1}} \mathrm{Tr}\bigg(\bold{R}_{1} \left( \frac{z}{\alpha} \bold{I}+ \frac{1}{\alpha} h_{2}h_{3}h_{4}h_{5} \bold{R}_{1}   \right)^{-1}
\end{aligned}
\vspace{-0.6cm}
\end{equation}

\begin{equation}
\label{scalable}
\begin{aligned}
&\times \!\! \left( \left(z- \frac{z}{\alpha} \right)  \bold{I} \!+\!
 \left( h^{\alpha}_{2}h^{\alpha}_{3}h^{\alpha}_{4}h^{\alpha}_{5} - \frac{1}{\alpha} h_{2}h_{3}h_{4}h_{5} \right) \bold{R}_{1}  \right)^{-1}\\
&\times \left(z\bold{I}+ h^{\alpha}_{2}h^{\alpha}_{3}h^{\alpha}_{4}h^{\alpha}_{5} \bold{R}_{1}   \right)^{-1}\bigg) >0,
\end{aligned}
\vspace{-0.2cm}
\end{equation}
where $h^{\alpha}_{i}$ is the solution under $\alpha h_1$. Since $\alpha h_1 > h_{1}$, from the conclusion in ii), we have $ \alpha h_{1}h^{\alpha}_{2}h^{\alpha}_{3} h^{\alpha}_{4}   h^{\alpha}_{5} > h_{1}h_{2}h_{3} h_{4}h_{5}$, which implies that $\alpha h^{\alpha}_{2}h^{\alpha}_{3}h^{\alpha}_{4}    h^{\alpha}_{5} > h_{2}h_{3} h_{4}h_{5}$ and~(\ref{scalable}) holds.

From the above, we know that $e(h_1)$ is a \textit{standard interference function}. Based on Theorem 2 in~\cite{yates1995framework}, for any initial value of $h^{0}_1>0$, the iterative equation $h^{t+1}_1=e\left( h^{t}_1 \right)$ converges to the unique positive solution.

\vspace{-0.3cm}




\end{IEEEproof}

\section{Proof of theorem 2}
\label{append2}
\begin{IEEEproof}
We consider the function
\begin{equation} 
\small
\label{r_x}
\overline{R}\left(x\right)=\overline{R}\left(x ,\! h_1,\!h_2,\!h_3,\!h_4,\!h_5\right).
 \vspace{-0.1cm}
\end{equation}
By combining~(\ref{eqt1}) to~(\ref{eqt5}), we obtain the partial derivatives with respect to $h_{i},~i=1,...,5$,
\begin{equation}
\small
 \frac{\partial  \overline{R}\left( x,\! h_1,\! h_2,\!h_3,\!h_4,\!h_5 \right)}{\partial h_{i}}   =  \frac{4M}{h_{i}}-  \frac{4M}{h_{i}} =0.
\end{equation}
Hence, we have
\vspace{-0.2cm}
\begin{equation}
\small
\begin{aligned}
 \frac{\partial\overline{R}\left( x \right)}{\partial x}& = \sum_{i=1}^{5}\! \frac{ \overline{R}\left( x,\! h_1,\! h_2, \!h_3, \!h_4, \!h_5 \right)}{\partial h_{i}} \frac{\partial{h_i}}{\partial x}\! +\! \frac{\overline{R}\left( x ,\! h_1,\! h_2,\!h_3,\!h_4,\!h_5\right)}{{\partial x}} \\
 &=\frac{1}{x}-\frac{1}{x^2n_{R_1}} \mathrm{Tr}\left(\left(\frac{1}{x}\bold{I}+h_{2}h_{3}h_{4}h_{5} \bold{R}_{1}  \right)^{-1}\right).
\end{aligned}
\vspace{-0.25cm}
\end{equation}
Since $\overline{R}\left( 0\right)=0$ and from~(\ref{r_std}), we have
\vspace{-0.15cm}
\begin{equation}
\small
\begin{aligned}
R\left( \sigma^2 \right) 
=\int_{0}^{\sigma^2} \frac{1}{z} - \frac{1}{z^2} \mathbb{E} \left[m_{\bold{B}}(\frac{1}{z})\right] dz 
\xrightarrow{n_{R_1}   \rightarrow \infty} \overline{R}\left( \sigma^2\right).
\end{aligned}
\end{equation}
This relation holds because the eigenvalues of $\bold{B}$ are bounded\cite{couillet2010deterministic}.
 Eventually, by substituting~(\ref{transform_t}) into~(\ref{r_x}) and the system of equations in~(\ref{de_group}), we obtain~(\ref{de_obj}) and complete the proof.     \hfill\IEEEQEDhere
\end{IEEEproof}

  \end{appendices}

\ifCLASSOPTIONcaptionsoff
  \newpage
\fi
\bibliographystyle{IEEEtran}
\bibliography{IEEEabrv,ref}

\end{document}